\documentstyle[11pt]{article}
\begin{document}
{\hfill\bf PUTP-95-28}
\vskip 2cm
\bigskip
\bigskip
\bigskip
{\Large\bf
	\centerline{Electromagnetic Annihilation Rates of $\chi_{c0}$ and 
$\chi_{c2}$ with}
\centerline{Both Relativistic and QCD Radiative Corrections}
\bigskip
\normalsize

\centerline{Han-Wen Huang$^{1,2}$~~~~Cong-Feng Qiao$^{1,2}$~~~~Kuang-Ta Chao$^{1,2}$}
\centerline{\sl $^1$ CCAST (World Laboratory), Beijing 100080, P.R.China} 
\centerline{\sl $^3$ Department of Physics, Peking University, Beijing 100871,
	   P.R.China}
\bigskip

\begin{abstract}
We estimate the electromagnetic decay rates of $\chi_{c0}\rightarrow\gamma\gamma$
and $\chi_{c2}\rightarrow\gamma\gamma$ by taking into 
account both relativistic
and QCD radiative corrections. The decay rates are derived in the 
Bethe-Salpeter formalism and the QCD radiative corrections are 
included in accordance with
factorization assumption. Using a QCD-inspired interquark potential,
we obtain relativistic BS wavefunctions of $\chi_{c0}$ and $\chi_{c2}$
by solving BS equations for the corresponding $^{2S+1}L_J$ states.  
Our numerical result for the ratio $R=\frac{\Gamma(\chi_{c0}\rightarrow
\gamma\gamma)}{\Gamma(\chi_{c2}\rightarrow\gamma\gamma)}$ is about $11-13$ 
which agrees with the update E760 experimental data. Explicit 
calculations show 
that the relativistic corrections due to spin-dependent interquark forces 
induced by gluon exchange enhance the ratio 
$R$ substantially and its value is insensitive to the choice of parameters
that characterize the interquark potential. Our expressions for the decay 
widths are identical with that obtained in
the NRQCD theory to the next-to-leading order in $v^2$ and $\alpha_s$. 
Moreover, we have determined two new coefficents in the nonperturbative matrix elements
for these decay widths.
\end{abstract}

\vfill\eject\pagestyle{plain}\setcounter{page}{1}

\section{Introduction}

~~~~~~~~Charmonium physics is in the boundary domain between perturbative and 
non-perturbative QCD. Charmonium decays may provide useful information on 
understanding the nature of interquark forces and decay mechanisms. Both QCD
radiative corrections and relativistic corrections are important for charmonium
decays, because for charmonium the strong coupling constant $\alpha_s\approx
0.3$ [defined in the $\overline{MS}$ scheme (the modified minimal subtraction 
scheme)] and the velocity squared of the quark in the meson rest frame $v^2
\approx 0.3$, both are not small.
Decay rates of heavy quarkonium in the nonrelativistic limit with QCD 
radiative
corrections have been studied (see, e.g.,ref.\cite{barbi,hag,mack,kwong}). 
However, the decay rates
of many processes are subject to substantial relativistic 
corrections \cite{kwong}. With 
this goal in mind, people have studied relativistic corrections 
to the decay rates of S-wave charmonium $\eta_c$,~$J/\psi$ and their radial 
excited states \cite{ktchao,keung,chiang}. These results show that relativistic effects are 
significant in the $c\bar{c}$ systems especially for the hadronic decays of 
$J/\psi$. In the present paper, we will investigate the
relativistic corrections to the electromagnetic decays of P-wave charmonium
states $\chi_{c0}\rightarrow\gamma\gamma$ and $\chi_{c2}\rightarrow\gamma
\gamma$.

The P-wave charmonium decays are interesting. Now their experimental results 
are quite uncertain. The Crystal Ball group (see \cite{mang,particle} and references therein) 
gives $\Gamma(\chi_{c0}\rightarrow\gamma
\gamma)=4.0\pm 2.8keV$. But for $\Gamma(\chi_{c2}\rightarrow\gamma\gamma)$,
its central value differs significantly among 
various experiments \cite{e760,cleo,tpc2},
and the ratio of the photonic widths for $\chi_{c0} and \chi_{c2}$ states 
measured by E760
is much larger than that measured by other two groups. Theoretically, in the
nonrelativistic limit, the ratio is $\frac{15}{4}$ \cite{barbi}, and it will increase to
about $7.4$ \cite{mang} if QCD radiative corrections are considered. Recently a rigorous 
factorization formula which is based on NRQCD has been developed for calculations 
of inclusive decay rates of heavy quarkonium. In this approach the decay widths
factor into a set of long distance matrix elements of NRQCD with each 
multiplied by a short distance
coefficient. To any given order of relative velocity $v$ of heavy quark and 
antiquark, the decay rates are determined by several nonperturbative factors
which can be evaluated using QCD lattice calculations or extracted by fitting
the data. The study of the photonic decays of $\chi_{c0}$ and $\chi_{c2}$
can also povide a determination for the nonperturbative factors in the decays of P-wave
quarkonium. 

In this paper, we will use the Bethe-Salpeter (BS) formalism \cite{bethe} to derive
the decay amplitudes and to calculate the decay widths of $\chi_{c0}\rightarrow
\gamma\gamma$ and $\chi_{c2}\rightarrow\gamma\gamma$. The meson will be treated
as a bound state consist of a pair of constitutent quark and antiquark (higher
Fock states such as $|Q\bar{Q}g>$ and $|Q\bar{Q}gg>$ are neglected because they
don't contribute to electromagnetic decays) and described by BS wavefunction
which satisfies the BS equation. A phenomenological QCD-inspired interquark
potential will be used to solve for the wavefunction and to calculate the decay
widths. Both relativistic and QCD radiative corrections to next-to-leading 
order will be considered based on the factorization assumption for the long
distance and short distance effects. The remainer of this paper is organized 
as follows. In Sec.2 we derive the 
reduced BS equation for any angular momentum state 
$^{2S+1}L_J$ of heavy mesons. In Sec.3 we give out the decay amplitudes of 
$\chi_J\rightarrow\gamma\gamma(J=0,2)$ and use the solved relativistic BS 
wavefunctions to calculate the numerical results of decay widths. A summary and 
discussion will be given in the last section.

\section{Reduced BS equations for any angular-momentum state $^{2S+1}L_J$  
of heavy mesons} 

~~~~~~~~Define the Bethe-Salpeter wavefunction, in general, for a $Q_1\bar{Q}_2$
bound state $|P>$ with overall mass M and momentum $P=(\sqrt{\vec{P}^2+
M^2},\vec{P})$
\begin{equation}
\chi(x_1,x_2)=<0|T\psi_1(x_1)\bar{\psi}_2(x_2)|P>,
\end{equation}
and transform it into momentum space
\begin{equation}
\chi_P(p)=e^{-iP\cdot X}\int d^4xe^{-ip\cdot x}\chi(x_1,x_2).
\end{equation}
Here $p_1(m_1)$ and $p_2(m_2)$ represent the momenta(masses) of quark and 
antiquark respectively,
$$
X=\eta_1x_1+\eta_2x_2,~~x=x_1-x_2,
$$
$$
P=p_1+p_2,~~p=\eta_2p_1-\eta_1p_2,
$$
where $\eta_i=\frac{m_i}{m_1+m_2}(i=1,2)$.
 
We begin with the bound state BS equation \cite{bethe} in momentum space 
\begin{equation}\label{bseq}
(\rlap/{p_1}-m_1)\chi_P(p)(\rlap/{p_2}+m_2)=\frac{i}{2\pi}\int d^4k
G(P,p-k)\chi_P(k),
\end{equation}
where $G(P,p-k)$ is the interaction kernel which dominates the 
interquark dynamics. In solving (\ref{bseq}), we will employ the instantaneous
approximation since for heavy quarks the interaction is dominated by 
instantaneous potentials. Meanwile, we will neglect negative energy projectors in
the quark propagators which are of even higher orders. Defining  
three dimensional BS wavefunction 
$$
\Phi_P(\vec{p})=\int dp^0\chi_P(p),
$$
we then get the reduced Salpeter equation for $\Phi_P(\vec{p})$
\begin{equation}
\label{bse}
(M-E_1-E_2)\Phi(\vec{p})=\Lambda^1_+\gamma_0\int d^3kG(P,\vec{p}-\vec{k})
\Phi(\vec{k})\gamma_0\Lambda^2_-.
\end{equation}
Here $G(P,\vec{p}-\vec{k})$ represents the instantaneous potential,  
$\Lambda_+(\Lambda_-)$ are the positive (negative) energy projector operators
for quark and antiquark respectively
$$
\Lambda^1_+=\frac{E_1+\gamma_0\vec{\gamma}\cdot\vec{p_1}+m\gamma_0}{2E_1}
$$
$$
\Lambda^2_-=\frac{E_2-\gamma_0\vec{\gamma}\cdot\vec{p_2}-m\gamma_0}{2E_2}
$$
$$
E_1=\sqrt{\vec{p_1}^2+m_1^2},~~E_2=\sqrt{\vec{p_2}^2+m_2^2}.
$$
We will follow a phenomenological approach by using QCD inspired inter-quark 
potentials, which are supported by both lattice QCD and heavy quark 
phenomenology, as the kernel in the BS equation. 
The potentials include a long-ranged 
confinement potential (Lorentz scalar) and a short-ranged one-gluon exchange 
potential (Lorentz vector) 
\begin{eqnarray}\label{pot}\nonumber
V(r)&=&V_S(r)+\gamma_{\mu}\otimes\gamma^{\mu}V_V(r),\\\nonumber
V_S(r)&=&\lambda r\frac{(1-e^{-\alpha r})}{\alpha r},\\
V_V(r)&=&-\frac{4\alpha_s(r)}{3r}e^{-\alpha r},
\end{eqnarray}
where the introduction of the factor $e^{-\alpha r}$ is to regulate the 
infrared divergence and also to incorporate the color screening effects of 
dynamical light quark pairs on the $Q\bar{Q}$ linear confinement potential. 
In momentum space the potentials become
\begin{eqnarray}\nonumber
G(\vec{p})&=&G_S(\vec{p})+\gamma_{\mu}\otimes\gamma^{\mu}G_V(\vec{p}),\\\nonumber
G_S(\vec{p})&=&-\frac{\lambda}{\alpha}\delta^3(\vec{p})+\frac{\lambda}{\pi^2}
\frac{1}{(\vec{p}^2+\alpha^2)^2},\\
G_V(\vec{p})&=&-\frac{2}{3\pi^2}\frac{\alpha_s(\vec{p})}{\vec{p}^2+\alpha^2},
\end{eqnarray}
where $\alpha_s(\vec{p})$ is the quark-gluon running coupling constant and is assumed to become a constant of $O(1)$ as $\vec{p}^2\rightarrow 0$
$$
\alpha_s(\vec{p})=\frac{12\pi}{27}\frac{1}{ln(a+\vec{p}^2/\Lambda^2_{QCD})}.
$$
The constants $\lambda,\alpha,a$ and $\Lambda_{QCD}$ are the parameters that 
characterize the potential. 

For any given angular-momentum state $^{2S+1}L_J$ 
of mesons, its three dimensional wavefunction in the rest frame of mesons takes the 
following two forms:\\ 
(i)S=0, then J=L,  
\begin{equation}
\label{wave1}
\Phi_{Lm}(\vec{p})=\Lambda^1_+\gamma_0(1+\gamma_0)
\gamma_5\gamma_0\Lambda^2_-Y_{Lm}(\hat{p})\phi(p);
\end{equation}
(ii)S=1, then J=L-1,L,L+1 for L$\ne$0 or J=1 for L=0,
\begin{equation}
\label{wave2}
\Phi_{JM}(\vec{p})=\sum_{l,m}<JM|1lLm>\Lambda^1_+\gamma_0(1+\gamma_0)
\gamma_l\Lambda^2_-\gamma_0Y_{Lm}(\hat{p})\phi(p)
\end{equation}
where $Y_{Lm}(\hat{p})$ is the spherical harmonic function and $<JM|1lLm>$ 
is the 
Clebsch--Gordan coefficient.
Substituting Eq.(\ref{wave1}) and (\ref{wave2}) in Eq.(\ref{bse}), 
one derives the equations for the scalar wavefunction $\phi(p)$\\
(i)S=0 
\begin{eqnarray}\label{bsec1}\nonumber
&&(M-E_1(p)-E_2(p))g_1(p)\phi(p)\\\nonumber
&=&-\frac{E_1(p)E_2(p)+m_1m_2+\vec{p}^2}{4E_1(p)E_2(p)}\int d^3k(G_S(\vec{p}-\vec{k})
-4G_V(\vec{p}-\vec{k}))g_1(k)P_L(cos\Theta)\phi(k)\\\nonumber
&-&\frac{E_1(p)m_2+E_2(p)m_1}{4E_1(p)E_2(p)}\int d^3k(G_s(\vec{p}-\vec{k})
+2G_V(\vec{p}-\vec{k}))g_2(k)P_L(cos\Theta)\phi(k)\\\nonumber
&+&\frac{E_1(p)+E_2(p)}{4E_1(p)E_2(p)}\int d^3kG_S(\vec{p}-\vec{k})\vec{p}\cdot\vec{k}
g_3(k)P_L(cos\Theta)\phi(k)\\
&+&\frac{m_1-m_2}{4E_1(p)E_2(p)}\int d^3k(G_S(\vec{p}-\vec{k})-2G_V(\vec{p}-\vec{k}))
\vec{p}\cdot\vec{k}g_4(k)P_L(cos\Theta)\phi(k)
\end{eqnarray}
where
\begin{eqnarray}\nonumber
g_1(p)&=&\frac{(E_1(p)+m_1)(E_2(p)+m_2)+\vec{p}^2}{4E_1(p)E_2(p)}\\\nonumber
g_2(p)&=&\frac{(E_1(p)+m_1)(E_2(p)+m_2)-\vec{p}^2}{4E_1(p)E_2(p)}\\\nonumber
g_3(p)&=&\frac{E_1(p)+m_1+E_2(p)+m_2}{4E_1(p)E_2(p)}\\\nonumber
g_4(p)&=&\frac{E_1(p)+m_1-E_2(p)-m_2}{4E_1(p)E_2(p)}\\\nonumber
E_1(p)&=&\sqrt{\vec{p}^2+m_1^2}\\\nonumber
E_2(p)&=&\sqrt{\vec{p}^2+m_2^2}
\end{eqnarray}
(ii)S=1
\begin{eqnarray}\label{bsec2}\nonumber
&&(M-E_1(p)-E_2(p))f_8(p)\phi(p)\\\nonumber
&=&\frac{1}{4E_1(p)E_2(p)}\{\int d^3k[(2G_V(\vec{p}-\vec{k})-G_S(\vec{p}-\vec{k}))
f_1(k)(m_1+m_2)\\\nonumber
&-&G_s(\vec{p}-\vec{k})f_2(k)(E_1(p)+E_2(p)]P_L(cos\Theta)\phi(k)\\\nonumber
&+&[\int d^3k(4G_V(\vec{p}-\vec{k})+G_S(\vec{p}-\vec{k}))f_8(k)(E_1(p)E_2(p)-m_1m_2
+\vec{p}^2)\\\nonumber
&+&(G_S(\vec{p}-\vec{k})-2G_V(\vec{p}-\vec{k}))f_7(k)(m_1E_2-m_2E_1)]
P_J(cos\Theta)\frac{k}{p}\phi(k)\\\nonumber
&+&\int d^3k[(2G_V(\vec{p}-\vec{k})-G_S(\vec{p}-\vec{k}))f_5(k)(m_1+m_2)\\
&-&G_S(\vec{p}-\vec{k})f_6(k)(E_1+E_2)]\vec{p}\cdot\vec{k}P_J(cos\Theta)
\frac{k}{p}\phi(k)\},
\end{eqnarray}
where
\begin{eqnarray}\nonumber
f_1(p)&=&\frac{1}{4E_1(p)E_2(p)}((E_1(p)+M_1)(E_2(p)+M_2)+\vec{p}^2)\\\nonumber
f_2(p)&=&\frac{1}{4E_1(p)E_2(p)}((E_1(p)+m_1)(E_2(p)+m_2)-\vec{p}^2)\\\nonumber
f_3(p)&=&f_4(p)=\frac{2(E_1(p)+m_1)}{4E_1(p)E_2(p)}\\\nonumber
f_5(p)&=&-f_6(p)=-\frac{2}{4E_1(p)E_2(p)}\\\nonumber
f_7(p)&=&\frac{1}{4E_1(p)E_2(p)}(E_1(p)+m_1-E_2(p)-m_2)\\\nonumber
f_8(p)&=&\frac{1}{4E_1(p)E_2(p)}(E_1(p)+m_1+E_2(p)+m_2).
\end{eqnarray}
The normalization condition $\int d^3pTr\{\Phi^+(\vec{p})\Phi(\vec{p})\}=\frac{2M}{(2\pi)^3}$ 
for the BS wavefunction $\phi(p)$ leads to
\begin{equation}\label{norma}
\int dpp^3\frac{(E_1(p)+m_1)(E_2(p)+m_2)}{4E_1E_2}\phi^2(p)=\frac{2M}{(4\pi)^3}.
\end{equation}
To the leading order in the nonrelativistic limit, Eqs.(\ref{wave1}) and (\ref{wave2}) 
are just the ordinary nonrelativistic Schr\"{o}dinger equation for orbital 
angular momentum $L$ with simply a spin-independent linear plus Coulomb potential. 
Solving equation (\ref{bsec1}) or (\ref{bsec2}), we can get the spectra and 
wavefunctions for any given angular-momentum state $^{2S+1}L_J$ of heavy 
mesons. With these wavefunctions we can calculate hadronic matrix elements 
of the processes involving corresponding states, and the relativistic 
corrections due to interquark dynamics are included automatically in them.
This approach is different from convential ones which start from 
Schr\"{o}dinger equation with all relativistic effects considered perturbatively. 

\section{Decay rates of $\Gamma(\chi_{c0}\rightarrow\gamma\gamma)$
and $\Gamma(\chi_{c2}\rightarrow\gamma\gamma)$}

~~~~~~~~Electromagnetic decays of $\chi_{c0}$ and $\chi_{c2}$ proceed via the annihilation 
of $c\bar{c}$ to two photons. Here only electromagnetic interactions
are considered, and color-octet components which contribute dominantly in 
hadronic decays of P-wave quarkonium do not contribute to electromagnetic decay widths, because 
final states are the photons which can not be produced via the annihilation of 
color-octet $Q\bar{Q}$ pair. So two photonic decays of $\chi_{cJ}$ for 
$J=0,~2$ can be 
well expressed in the BS formalism and relativistic corrections are 
incorporated systematicly in the decay rates. In the BS formalism the 
annihilation matrix elements can be written as follows
\begin{equation}
<0|\bar{Q}IQ|P>=\int d^4pTr[I(p,P)\chi_P(p)],
\end{equation}
where $I(p,P)$ is the interaction vertex of the $Q\bar{Q}$ with other fields
(e.g., the photons or gluons) which, in general. may also depend on the 
variable $q^0$ (the time-component of the relative momentum). If $I(p,P)$ is 
independent of $q^0$ (e.g., if quarks are on their mass-shells in the 
annihilation), the equation can be written as
\begin{equation}
<0|\bar{Q}IQ|P>=\int d^3pTr[I(\vec{p},P)\Phi_P(\vec{p})],
\end{equation}

For process $\chi_{c0}\rightarrow\gamma\gamma$ or $\chi_{c2}\rightarrow
\gamma\gamma$ with the momenta and polarizations of photons $k_1,\epsilon_1$
and $k_2,\epsilon_2$, the decay amplitude can be written as
\begin{equation}
T=<0|\bar{c}\Gamma_{\mu\nu}c|\chi_{cJ}>\epsilon^{\mu}_1\epsilon^{\nu}_2
\end{equation}
for J=0,2, where $p_1(p_2)$ is the charm quark(antiquark) momentum, and
$$
\Gamma_{\mu\nu}=e^2[\gamma_{\nu}\frac{1}{\rlap/{p_1}-\rlap/{k_1}-m}
\gamma_{\mu}+\gamma_{\mu}\frac{1}{\rlap/{k_1}-\rlap/{p_2}-m}\gamma_{\nu}]
$$
Since $p^0_1+p^0_2=M$, as usual we take
\begin{equation}\label{on-shell}
p^0_1=p^0_2=\frac{M}{2}.
\end{equation}
Therefore, the amplitude $T$ becomes independent of $p^0$. In terms of $T$, the 
decay rates can be written as 
\begin{equation}\label{decay}
\Gamma(\chi_{cJ}\rightarrow\gamma\gamma)=\frac{1}{2!}\sum_{spin}\sum_{polar}
\int|T|^2d\Omega 
\end{equation}
for $J=0,2$, where the factor $1/2!$ is needed because $N!$ same graphs 
appear for N-photon
final states. The photon polariztion is summed over in the Feynman gauge,
$$
\sum_{helicity}\varepsilon^{\mu}(k_1)\varepsilon^{\nu*}(k_1)=-g^{\mu\nu},
$$
Substituting BS wavefunction (\ref{wave2}) into (\ref{decay}), 
after summing over final states and averaging over initial states, we get
\begin{equation}\label{wid1}
\Gamma(\chi_{c0}\rightarrow\gamma\gamma)=24e^4_Q\alpha^2(c_1+3c_2+2c_3)^2
\end{equation}
\begin{equation}\label{wid2}
\Gamma(\chi_{c2}\rightarrow\gamma\gamma)=\frac{12e^4_Q\alpha^2}{5}
(c_1^2-2c_1c_3+7c_3^2)
\end{equation}
where
\begin{eqnarray}\nonumber
c_1&=&\int d^3p\frac{1}{(\vec{p}-\vec{k})^2+m^2}\{[-E^2-mE-\frac{\vec{p}^2}{2}
+\frac{3(\vec{p}\cdot\hat{k})^2}{2}]\vec{p}\cdot\vec{k}\\\nonumber
&&+[-\frac{\vec{p}^4}{4}+\frac{3\vec{p}^2(\vec{p}\cdot\hat{k})}{2}
-\frac{5(\vec{p}\cdot\hat{k})^4}{4}]\}\frac{\phi(p)}{p}\\\nonumber
c_2&=&\int d^3p\frac{1}{(\vec{p}-\vec{k})^2+m^2}\{\frac{\vec{p}\cdot\vec{k}}
{2}[\vec{p}^2-(\vec{p}\cdot\hat{k})^2]+\frac{1}{4}[\vec{p}^2-(\vec{p}\cdot
\hat{k})^2]^2\}\frac{\phi(p)}{p}\\\nonumber
c_3&=&\int d^3p\frac{1}{(\vec{p}-\vec{k})^2+m^2}\{\frac{-E^2-mE}
{2}[\vec{p}^2-(\vec{p}\cdot\hat{k})^2]+\frac{1}{4}[\vec{p}^2-(\vec{p}\cdot
\hat{k})^2]^2\}\frac{\phi(p)}{p}\\\nonumber
\end{eqnarray}
In the nonrelativistic limit, (\ref{wid1}) and (\ref{wid2}) reduce to
$$
\Gamma(\chi_{c0}\rightarrow\gamma\gamma)=\frac{24e^4_Q\alpha^2}{m^4}
|\int d^3pp\phi_{\chi_{c0}}(p)|^2
$$
$$
\Gamma(\chi_{c2}\rightarrow\gamma\gamma)=\frac{32e^4_Q\alpha^2}{5m^4}
|\int d^3pp\phi_{\chi_{c2}}(p)|^2
$$
Using the Fourier transformation of wavefunctios 
$$
\int d^3pp\phi_{\chi_{cJ}}(p)=\frac{3}{\sqrt{8}}R_{\chi_{cJ}}^{\prime}(0)
$$
we derive the well known results in coordinate space, which
is consistent with that given in \cite{barbi}
\begin{equation}
\Gamma(\chi_{c0}\rightarrow\gamma\gamma)=\frac{27e^4_Q\alpha^2}{m^4}
|R^{\prime}_{\chi_{c0}}(0)|^2
\end{equation}
\begin{equation}
\Gamma(\chi_{c}\rightarrow\gamma\gamma)=\frac{36e^4_Q\alpha^2}{5m^4}
|R^{\prime}_{\chi_{c2}}(0)|^2,
\end{equation}
where $R^{\prime}_{\chi_cJ}(0)$ is the derivative of radial 
wavefunction at the origin,
and in the nonrelativistic limit, $R^{\prime}_{\chi_{c0}}(0) = 
R^{\prime}_{\chi_{c2}}(0)$,
 due to the heavy quark spin symmetry.

Recently, in the framework of NRQCD the factorization formulas for the long 
distance and short distance effects were found to involve a double expansion 
in the quark relative velocity $v$ and in the QCD coupling constant $\alpha_s$ 
\cite{BBL,BBL1}. To next-to-leading order in both $v^2$ and $\alpha_s$, as an approximation, we may write
\begin{equation}\label{wids1}
\Gamma(\chi_{c0}\rightarrow\gamma\gamma)=24e^4_Q\alpha^2(c_1+3c_2+2c_3)^2
[1+\frac{\alpha_s}{\pi}(\frac{\pi^2}{3}-\frac{28}{9})]
\end{equation}
\begin{equation}\label{wids2}
\Gamma(\chi_{c2}\rightarrow\gamma\gamma)=\frac{12e^4_Q\alpha^2}{5}
(c_1^2-2c_1c_3+7c_3^2)
(1-\frac{\alpha_s}{\pi}\frac{16}{3})
\end{equation}
where we have used QCD radiative corrections given in \cite{barbi1}. 
We must emphasize 
that above factorization formula are  correct only to next-to-leading order in 
$v^2$ and $\alpha_s$. If higher order effects are involved, the decay widths can 
not be factored into a integral of wavefunction and a coefficient that can be written 
as a series of $\alpha_s$. NRQCD has applied a more general factorization formula for 
quarkonium decay rates, which will be discussed in detail later.

For the heavy quarkonium $c\bar{c}$ systems, $m_1=m_2=m_c$, Eqs. (\ref{bsec1}) and (\ref{bsec2}) 
become much simpler. We take the following parameters 
which appear in potential (\ref{pot}),
$$
m_c=1.5Gev,~~\lambda=0.23Gev^2,~~\Lambda_{QCD}=0.18Gev,
$$
$$
\alpha=0.06Gev,~~a=e=2.7183.
$$
With these values the mass spectrum of charmonium are found to fit the 
data well. 
In Fig.1 and Fig.2 the solved scalar wavefunctions both in momentum and coordinate space 
for P-wave triplet $\chi_{cJ}$ states are shown 
and we can see explicitly the differences between wave functions for
 $J=0,~2$  
but they are same in the nonrelativistic limit.
Substituting $\phi_{\chi_{c0}}(p)$ and $\phi_{\chi_{c2}}(p)$ into 
(\ref{wids1}) and (\ref{wids2}), we get 
$$
\Gamma(\chi_0\rightarrow\gamma\gamma)=5.32keV,
$$
$$
\Gamma(\chi_2\rightarrow\gamma\gamma)=0.44keV
$$
their ratio is
\begin{equation}\label{ratio1}
R=\frac{\Gamma(\chi_0\rightarrow\gamma\gamma)}
{\Gamma(\chi_2\rightarrow\gamma\gamma)}=12.1.
\end{equation}
Our results are satisfactory. as compared with the Particle Data Group 
experimental values \cite{particle} $\Gamma(\chi_{c0}\rightarrow\gamma\gamma)=
5.6\pm 3.2keV$, and $\Gamma(\chi_{c2}\rightarrow\gamma\gamma)=0.32\pm 0.1keV$. 
Here in above calculations the value of $\alpha_s(m_c)$ in the QCD radiative
correction factor in (\ref{wids1}) and (\ref{wids2}) is chosen to be 0.29 \cite{kwong}, which
is also consistent with our determination from the ratio of 
$B(J/\psi\rightarrow 3g)$ to $B(J/\psi\rightarrow e^+e^-)$ \cite{ktchao} .

Moreover, in order to see the sensitivity of the decay widths to the parameters,
especially the charm quark mass, we use other two sets of parameters
$$
m_c=1.4Gev,~~\lambda=0.24Gev^2, 
$$
$$
m_c=1.6Gev,~~\lambda=0.22Gev^2, 
$$
with other parameters keeping unchanged (the heavy quarkonia mass spectra are not 
sensitive to $a,~\alpha$ for $\alpha\le 0.06GeV$), By the same 
procedure, we obtain
$$
\Gamma(\chi_0\rightarrow\gamma\gamma)=5.82(4.85)keV,
$$
$$
\Gamma(\chi_2\rightarrow\gamma\gamma)=0.50(0.39)keV,
$$
and the ratio 
\begin{equation}\label{ratio2}
R=\frac{\Gamma(\chi_0\rightarrow\gamma\gamma)}
{\Gamma(\chi_2\rightarrow\gamma\gamma)}=11.8(12.5)
\end{equation}
for $m_c=1.4(1.6)GeV$.

We find the widths are decreased with the decreasing of $\lambda$ This 
is obvious since 
the wavefunction in coordinate space will become broader when the slope of linear potential 
is decreased and the corresponding wavefunction in momentum space will become narrower,  
so the effective couplings decay become smaller.   
It is interesting to note that the ratio of two photonic decay width of $\chi_{c0}$ and $\chi_{c2}$ is almost unchanged
and is insensitive to the choice of parameters. 

Finally we discuss the relation between our approach and the NRQCD theory.
Recently, a general
factorization formula which is based on nonrelativistic QCD (NRQCD) 
has been developed for studying the inclusive cross sections of production and 
decay of heavy quarkonium. In this formalism the quarkonium decay rates
can be written as a sum of a set of matrix elements to any given order
in $v^2$, with each matrix element multiplied by a coefficient which can be
calculated in perturbative QCD. This approach has been proved
successful in the application of some processes involving heavy quarkonium \cite{pwave,prod}.
In NRQCD, the electromagnetic decay rates of $\chi_{c0}$
and $\chi_{02}$ to next-to-leading order in $v^2$ can be written as
\begin{equation}
\Gamma(\chi_0\rightarrow\gamma\gamma)=\frac{2Imf_{EM}(^3p_0)}{m^4}
<\chi_{c0}|{\cal O}_{EM}(^3P_0)|\chi_{c0}>+\frac{2Img_{EM}(^3p_0)}{m^6}
<\chi_{c0}|{\cal O}_{EM}(^3P_0)|\chi_{c0}>
\end{equation}
\begin{equation}
\Gamma(\chi_2\rightarrow\gamma\gamma)=\frac{2Imf_{EM}(^3p_2)}{m^4}
<\chi_{c2}|{\cal O}_{EM}(^3P_2)|\chi_{c2}>+\frac{2Img_{EM}(^3p_2)}{m^6}
<\chi_{c2}|{\cal O}_{EM}(^3P_2)|\chi_{c2}>
\end{equation}
where
\begin{eqnarray}\label{ope}\nonumber
{\cal O}_{EM}(^3P_0)&=&
\frac{1}{3}\psi^+(-\frac{i}{2}\stackrel{\leftrightarrow}{\bf D})
\cdot\vec{\sigma}\chi|0><0|\chi^+
(-\frac{i}{2}\stackrel{\leftrightarrow}{\bf D})
\cdot\vec{\sigma}\psi\\\nonumber
{\cal O}_{EM}(^3P_2)&=&
\psi^+(-\frac{i}{2}\stackrel{\leftrightarrow}{D})^{(i}
\sigma^{j)}\chi|0><0|\chi^+
(-\frac{i}{2}\stackrel{\leftrightarrow}{D})^{(i}
\sigma^{j)}\psi\\\nonumber
{\cal G}_{EM}(^3P_0)&=&
\frac{1}{2}[\frac{1}{3}\psi^+
(-\frac{i}{2}\stackrel{\leftrightarrow}{\bf D})^2
(-\frac{i}{2}\stackrel{\leftrightarrow}{\bf D})
\cdot\vec{\sigma}\chi|0><0|\chi^+
(-\frac{i}{2}\stackrel{\leftrightarrow}{\bf D})
\cdot\vec{\sigma}\psi+h.c]\\
{\cal G}_{EM}(^3P_0)&=&
\frac{1}{2}[\psi^+
(-\frac{i}{2}\stackrel{\leftrightarrow}{\bf D})^2
(-\frac{i}{2}\stackrel{\leftrightarrow}{D})^{(i}
\sigma^{j)}\chi|0><0|\chi^+
(-\frac{i}{2}\stackrel{\leftrightarrow}{D})^{(i}
\sigma^{j)}\psi+h.c]
\end{eqnarray}
where $\vec{D}$ is the space component of covariant derivate $D^{\mu}$,
$\psi$ and $\chi$ are two component operators of quark and antiquark
respectively. If identifying the quark operator expectation values with the 
deriatives of wavefunctions at the origin, the decay widths
can be written as
\begin{equation}\label{rate1}
\Gamma(\chi_0\rightarrow\gamma\gamma)=\frac{9Imf_{EM}(^3p_0)}{\pi m^4}
|R^{\prime}_{\chi_0}(0)|^2+\frac{15Img_{EM}(^3p_0)}{\pi m^6}
Re(R^{(3)}_{\chi_0}(0)R^{\prime}_{\chi_0}(0))
\end{equation}
\begin{equation}\label{rate2}
\Gamma(\chi_2\rightarrow\gamma\gamma)=\frac{9Imf_{EM}(^3p_2)}{\pi m^4}
|R^{\prime}_{\chi_2}(0)|^2+\frac{15Img_{EM}(^3p_2)}{\pi m^6}
Re(R^{(3)}_{\chi_2}(0)R^{\prime}_{\chi_2}(0))
\end{equation}

In comparision, we take the on-shell condition $p^0_1=p^0_2=E$ instead of 
(\ref{on-shell}) and expand the annihilation amplitudes in (\ref{wid1}) and (\ref{wid2}) in terms of
$\vec{p}^2/m^2$. The leading order contribution
of P-wave decay comes from terms linear in $\vec{p}$ and  in order to take into account  
their higher order effects we retain those terms proportional to 
the third powers of $\vec{p}$ then have
\begin{equation}
\label{per0}
\Gamma(\chi_0\rightarrow\gamma\gamma)=\frac{3e^4_Q\alpha^2}{m^4}
|\int d^3pp(1-\frac{\vec{p}^2}{6m^2})\phi_{Sch}(p)|^2
[1+\frac{\alpha_s}{\pi}(\frac{\pi^2}{3}-\frac{28}{9})]
\end{equation}
\begin{equation}
\label{per2}
\Gamma(\chi_2\rightarrow\gamma\gamma)=\frac{4e^4_Q\alpha^2}{5m^4}
|\int d^3pp\phi_{Sch}(p)|^2(1-\frac{\alpha_s}{\pi}\frac{16}{3}))
\end{equation}
where the stanard Schr\"{o}dinger wavefunction (with relativistic corrections) 
$\phi_{Sch}(p)$ is related to $\phi_(p)$ through the normalization condition (\ref{norma})
$$
\phi_{Sch}(p)=\frac{1}{\sqrt{M}}(\frac{m+E}{E})\phi(p),
$$
$$
(2\pi)^3\int dpp^2|\phi_{Sch}(p)|^2=1.
$$
Using formula
$$
\int d^3pp\phi_{Sch}(p)=3R^{\prime}_{Sch}(0),
$$
$$
\int d^3pp^3\phi_{Sch}(p)=5R^{(3)}_{Sch}(0),
$$
the expressions of (\ref{per0}) and (\ref{per2}) are transfered into coordinate space and
comparing with that derived from NRQCD (\ref{rate1}) and (\ref{rate2}), we can easily determine the coefficients 
\begin{eqnarray}\label{e1}\nonumber
Imf_{EM}(^3P_0)&=&3\pi e^4_Q\alpha^2[1+\frac{\alpha_s}{\pi}(\frac{\pi^2}{3}
-\frac{28}{9})]\\\nonumber
Img_{EM}(^3P_0)&=&-\pi e^4_Q\alpha^2\\\nonumber
Imf_{EM}(^3P_2)&=&\frac{4\pi e^4_Q\alpha^2}{5}(1-\frac{\alpha_s}{\pi}\frac{16}{3})\\
Img_{EM}(^3P_2)&=&0
\end{eqnarray}
Here we only consider the QCD radiative corrections to 
leading order coefficients 
$Imf_{EM}(^3P_0)$ and $Imf_{EM}(^3P_2)$ which are equal to the results 
derived in \cite{BBL}. 
More, we have determined two new coefficents, i.e, the second one and fourth
one in (\ref{e1}). These
two matrix elements have a suppression factor of $v^2$ so we need not to take into account 
higher order corrections to their coefficients any more.

\section{Summary and Discussion}

~~~~~~~~In this paper we provide an estimate for the photonic decays of P-wave
charmonium. It is clear from above calculations that comparing with  
nonrelativistic results the relativistic effects enhance the ratio $R$ 
substantially. We know that there are two sources of relativistic corrections: 
1)the correction due to relativistic kinematics which appears explicitly in the 
decay amplitudes; 2)the correction due to inter-quark dynamical effects (e.g. 
the well known Breit-Fermi interactions), which mainly causes the correction 
to the bound state wave functions. From the expressions (\ref{per0}) and (\ref{per2}) of decay rates
which have been expanded to the firdt order of $\frac{\vec{p}^2}{m^2}$, one might expect
that the ratio $R$ would become smaller after taking relativistic corrections into account,
because the coefficient of the term $\frac{\vec{p}^2}{m^2}$ in (\ref{per0}) is smaller 
than that in (\ref{per2}). 
However, mainly due to the attractive spin-orbital force induced by one gluon exchange
for the $0^{++}$ meson, the $\chi_0$ wavefunction becomes narrower than $\chi_2$ 
wavefunction in coordinate space, and therefore the derivative of wavefunction at the orgin become larger for
$\chi_0$ than that for $\chi_2$. As a result, the dynamic relativistic effect on $R$ is
in the opposite direction and can be even larger. The overall relativistic
correction to $R$ is found to be positive, and our result is in agreement with
the E760 data and disagree with the values mesured by CLEO and TPC2.

Our expressions for the decay widths are identical with that derived from the rigorous 
factorization formula to next-to-leading order in $v^2$ and in $\alpha_s$. 
Moreover we have determined two new coefficents in the nonperturbative matrix elements
for these decay widths.For a 
more accurate estimate, higher order corrections both in $v^2$ and in $\alpha_s$ should be
taken into account. For electromagnetic decays, in general we can estimate them 
within the $|Q\bar{Q}>$ sector and avoid the difficult problem due to 
the effects of high Fock states such as $|Q\bar{Q}g>$ and $|Q\bar{Q}gg>$. But
we must notice that if higher order matrix elements are included, the decay
widths can not be factored in the way like (\ref{rate1}) and (\ref{rate2}) because 
the higher order coefficients are
different for each nonperturbative factor. 

We have solved the BS equation for the bound-state wave functions with QCD 
inspired interquark potentials (linear confinement potential plus one gluon
exchange potential) as the BS kernel. With some popular parameters for the
potentials we obtained the wave functions and used them to calculate the decay
widths. From (\ref{ratio1}) and (\ref{ratio2}) it can be seen that three 
different wave functions lead
to somewhat different photonic decay widths but give very close values for $R$. 
This might indicate that our estimate of $R$ is insensitive to the quark mass and potential parameters,
and therefore could be a rather reliable result, despite 
the uncertainty in the estimate of the dynamical relativistic 
effects. We hope the lattice simulations will give more reliable estimates for 
these decays within the framework of NRQCD, and can be compared with our results.
\vfill\eject

\vfill\eject

\newpage


Fig.1  Wave functions $\phi_{Sch}(p)$ (normalized in momentum space)
of $\chi_{c0}$ (solid line), $\chi_{c1}$ (dashed line) and $\chi_{c2}$
(dashed and dot line) by solving BS equations with $m_c=1.5GeV$

Fig.2  Wave functions $R_{Sch}(x)$ (normalized in coordinate space)
of $\chi_{c0}$ (solid line), $\chi_{c1}$ (dashed line) and $\chi_{c2}$
(dashed and dot line) by solving BS equations with $m_c=1.5GeV$


\begin{thebibliography}{99}
\bibitem{barbi}{R. Barbieri, R. Gatto, and R. K\"{o}gerler,
	       {\it Phys. Lett.} {\bf 60B}, 183(1976); R. Barbieri, M. Caffo,
	       and E. Remiddi, {\it Phys. Lett.}{\bf 61B}, 465(1976);
	       R.Barbieri $et al$. {\it Nucl. Phys.} {\bf B162}, 220(1980).}
\bibitem{hag}{K. Hagiwara, C. B. Kim, and T. Yoshino, {\it Nucl. Phys.}
   {\bf B177}, 461(1981).}
\bibitem{mack}{P. B. Mackenzie and G. P. Lepage, {\it Phys. Rev. Lett.}
      {\bf 47}, 1244(1981).}
\bibitem{kwong}{W. Kwong, P. B. Mackenzie, R. Rosenfeld, and J. L. Rosner,
   {\it Phys. Rev.}{\bf D37}, 3210(1988).}
\bibitem{ktchao}{K. T. Chao, H. W. Huang, J. H. Liu, Y. Q. Liu, and J. Tang,
in Proceedings of {\it the International Conference on Quark Confinement and
the Hadron Spectrum}, Come, Italy, June 1994, edited by N. Brambilla and G. M.
Prosperi (World Scientific, Singapore)p.30 6.; K. T. Chao, H. W. Huang, and
Y. Q. Liu, hep-ph/9503201, to be published in {\it Phys. Rev.}{\bf D.}}
\bibitem{keung}{W. Y. Keung and I. J. Muzinich, {\it Phys. Rev.}{\bf D27},
1518(1983).}
\bibitem{chiang}{H. C. Chiang, J. Hufner and H. J. Pirner,{\it Phys. Lett.}
{\bf B324}, 482(1994).}
\bibitem{mang}{M. L. Mangano, and A. Petrelli,
		CERN preprint CERN-TH/95-67 (hep-ph/9503465).}
\bibitem{e760}{E760 Collaboration (T. A. Armstrong {\it et al.}),
	     {\it Nucl. Phys.}{\bf B373}, 35(1992); {\it ibid.}{\it
	     Phys. Rev. Lett.}{\bf 70}, 2988 (1993).}
\bibitem{cleo}{CLEO Collaboration (J. Dominick $et~al.$), {\it Phys. Rev.}
{\bf D50}, 4265(1994).}
\bibitem{tpc2}{TPC/Two-Gamma Collaboration, {\it Phys. Lett}. {\bf 302B},
345(1993).}
\bibitem{bethe}{E. E. Salpeter and H. A. Bethe, {\it Phys. Rev.} {\bf 84}, 1232(1951;
E. E. Salpeter, {\it Phys. Rev.} {\bf 87}, 328(1952).}
\bibitem{BBL}{G. T. Bodwin, E. Braaten, G. P. Lepage,
		{\it Phys. Rev.} {\bf D51}, 1125 (1995).}
\bibitem{BBL1}{E. Braaten, NUHEP-TH-94-22 (hep-ph/9409286).}
\bibitem{barbi1}{R. Barbieri, R. Gatto, M. Caffo, and E. Remiddi, 
{\it Nucl, Phys.} {\bf B192}, 61(1981).} 
\bibitem{particle}{Particle Data Group, L. Montanet $et~al$., {\it Phys.
Rev.}{\bf D50(3-I)},1171(1994).}
\bibitem{pwave}{G. T. Bodwin, E. Braaten, and G. P. Lepage,
       {\it Phys. Rev.} {\bf 46D}, 1914 (1992).}
\bibitem{prod}{E. Braaten, T. C. Yuan, {\it Phys. Rev. Lett.}{\bf 71},
  1673(1993); E. Braaten, S. Fleming, {\it Phys. Rev. Lett.}{\bf 74}, 3327 
	      (1995).} 
\end{thebibliography}
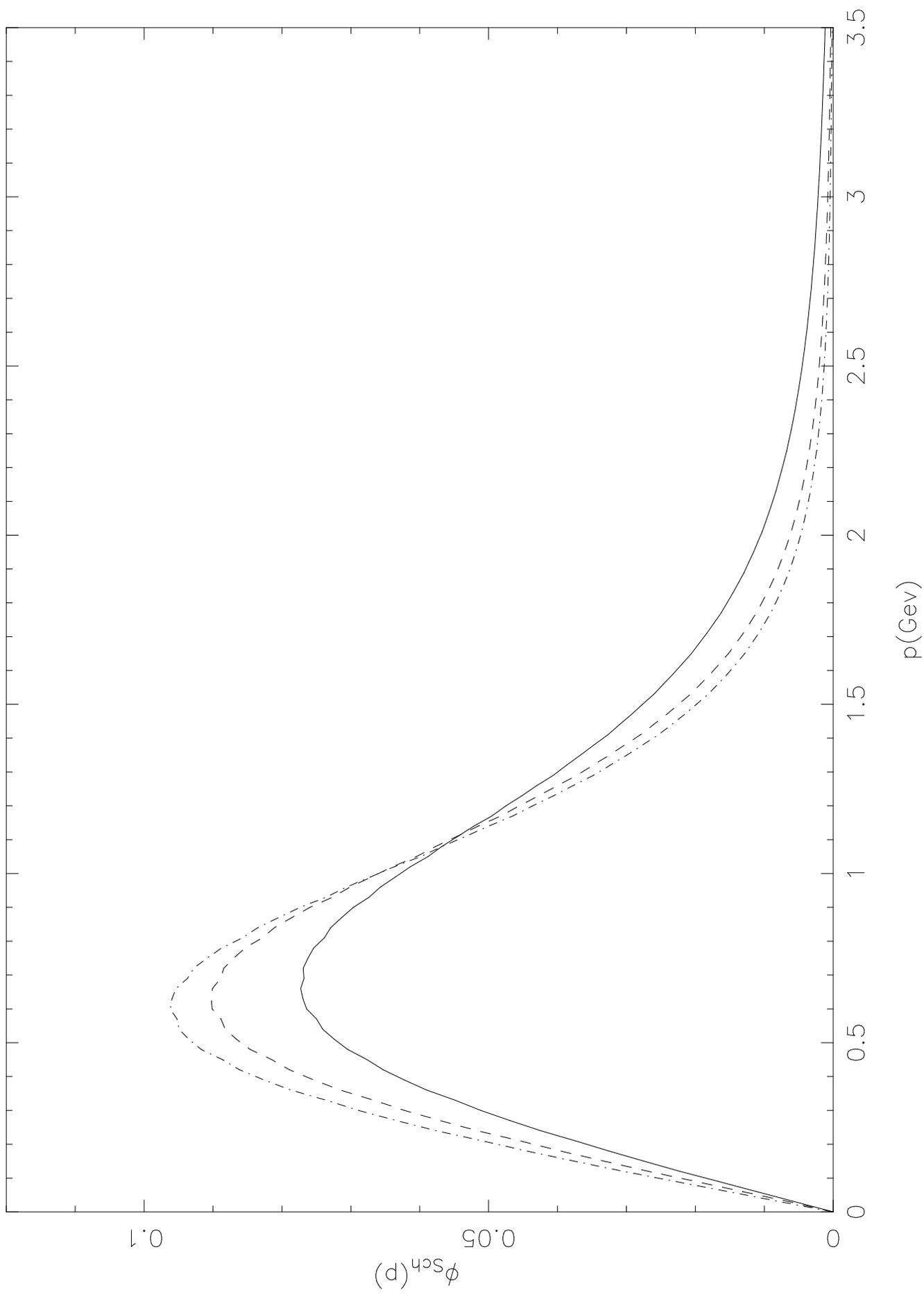
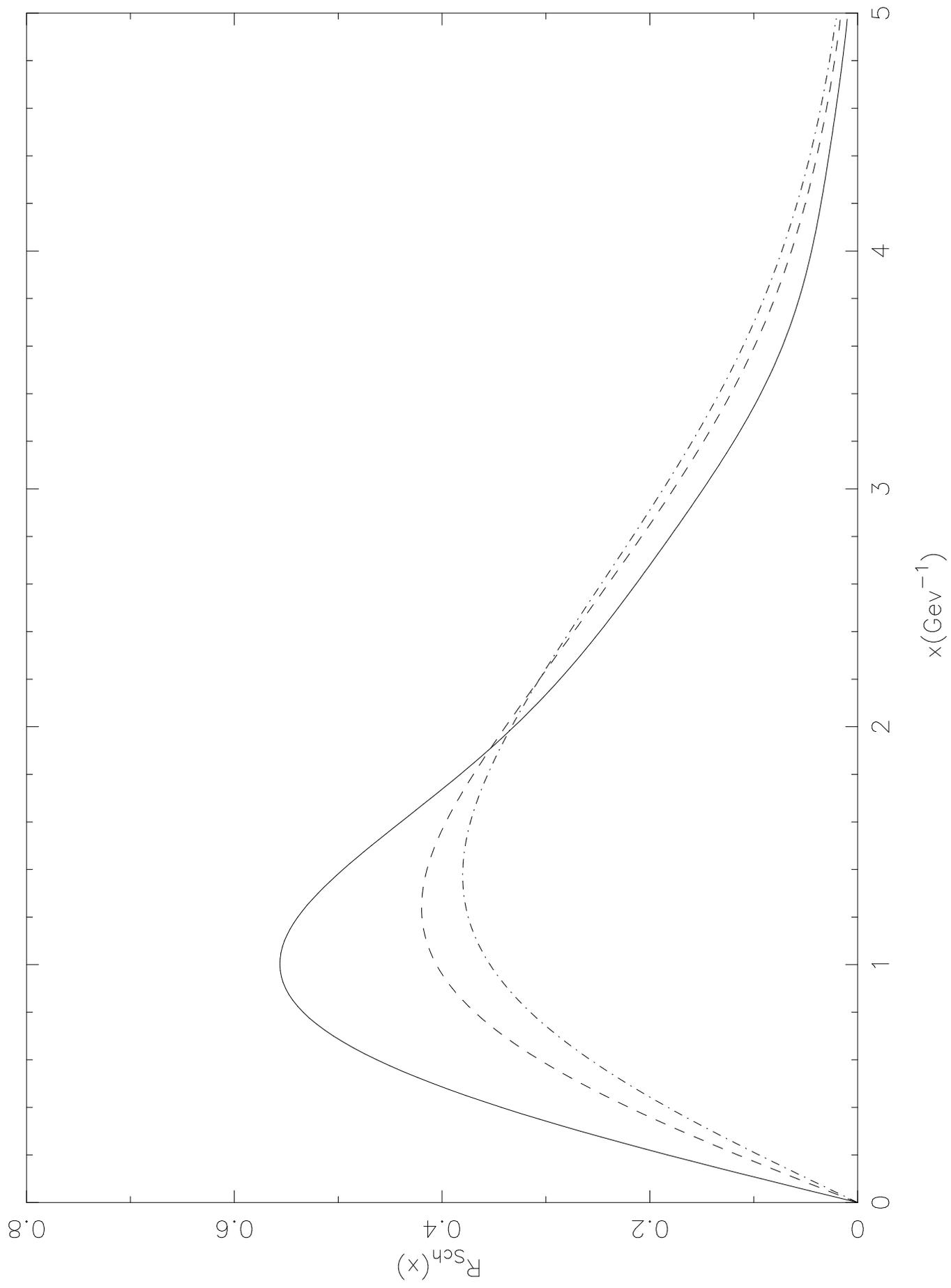
\end{document}